# INTENSIVE ELECTRON ANTINEUTRINO SOURCE WITH WELL DEFINED HARD SPECTRUM ON THE BASE OF NUCLEAR REACTOR AND 8-LITHIUM TRANSFER. THE PROMISING EXPERIMENT FOR STERILE NEUTRINOS SEARCH

## V. I. Lyashuk [1,2]

[1] *Institute for Nuclear Research, Russian Academy of Sciences, Moscow, Russia*
[2] *National Research Center "Kurchatov Institute", Moscow, Russia*

**Abstract:** The new concept of $\bar{\nu}_e$-source for future short-baseline experiments is discussed. The source ensures: 1) well defined hard antineutrino flux; 2) the rate of counts more than $\sim(10^2$-$10^3)$ per day in the detector volume about m$^3$; 3) low level of count errors - $\lesssim 1\%$. The proposed source is based on (n,γ)-activation of $^7$Li near the reactor active zone and transport of the fast $\beta^-$-decaying $^8$Li isotope toward a remote neutrino detector and back in the closed loop. The proposed experiment allows to detect ($\bar{\nu}_e$,p)-interaction with high precision and directed to search of sterile neutrinos with $\Delta m^2 \sim 1$ eV$^2$. The results of simulation for (3+1) and (3+2) model indicate the space regions for search of $\bar{\nu}_e$-disappearance.

In spite of the apparent superiority on neutrino flux the nuclear reactors has a disadvantage: too-small hardness of $\bar{\nu}_e$-spectrum that is extremely negative. For the considered here reactor antineutrino energy the neutrino cross section is proportional to it's energy squared - $\sigma_\nu \sim E_\nu^2$. Antineutrinos $\bar{\nu}_e$ emitted at $\beta^-$-decay of fission fragments in a nuclear reactor have rapidly decreasing spectrum and energy $E_{\bar{\nu}} \leq 10$ MeV. The errors of the spectrum is known with ~5% errors and complicate the analysis. The detected bump in reactor $\bar{\nu}_e$-spectrum [1] is one more evidence.

The disadvantage of rapidly dropping spectrum can be filled having realized the idea [2] to use a high-purified isotope $^7$Li for construction of lithium blanket (or converter) around the active zone (AZ) of a reactor [3]. A short-lived isotope $^8$Li ($T_{1/2} = 0.84$ s) is created under AZ neutrons flux in reaction $^7$Li(n,γ)$^8$Li and at $\beta^-$-decay it emits hard antineutrinos of a well determined spectrum with the maximal energy $E_{\bar{\nu}}^{max} \simeq 13.0$ MeV and mean one $\bar{E}_{\bar{\nu}} \simeq 6.5$ MeV.

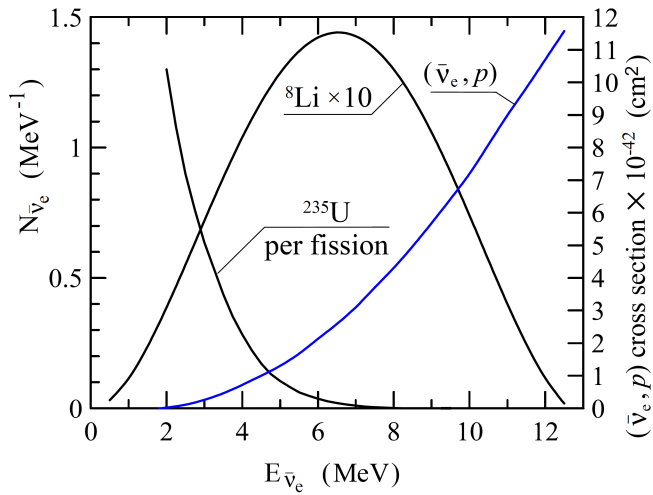

The $^{235}$U neutrino spectrum (the main fuel component) is presented in Fig. 1 in comparison with $^8$Li neutrino spectrum [4, 5]. The advantages of hard $\bar{\nu}_e$-spectrum of $^8$Li is clear on the example of sharp rise for cross section of inverse beta decay reaction $\bar{\nu}_e + p \rightarrow n + e^+$.

FIG. 1. Spectrum of antineutrinos from $\beta^-$-decay of $^8$Li [5] and $^{235}$U fission fragments [4] (see left axis). Cross section of $(\bar{\nu}_e, p)$−reaction is on the right axis [6].

Lithium blanket around the AZ (acting as a converter of reactor neutrons to antineutrinos) is the most simple scheme of lithium antineutrino source with steady spectrum source [3,7-9]. As a result the total $\bar{\nu}_e$-spectrum from AZ and from decays of $^8$Li isotope becomes considerably harder in comparison with the pure reactor neutrino spectrum, which errors which strongly rise at the energy above ~(5-6) MeV [10, 9, 11]. Note that reactor antineutrino spectrum is specified also with instability fuel composition ($^{235}$U, $^{239}$Pu, $^{238}$U, $^{241}$Pu) in operation period.

Let us define the productivity factor of the blanket $k$ (or coefficient of blanket efficiency) as number of $^8$Li nuclei produced in the lithium blanket per one fission in AZ. The hardness of the total spectrum will more larger as productivity factor $k$ will be more higher.

For our purpose (creation of the neutrino source of significantly larger hardness than possible to obtain by above mentioned simple scheme) let us introduce the definition of the generalized hardness for total neutrino spectrum [12,13]. Let $F_{\text{Li}}(\vec{r})$ and $F_{\text{AZ}}(\vec{r})$ - densities of lithium $\bar{\nu}_e$-fluxes from the blanket and from AZ, $\bar{n}_\nu \simeq 6.14$ - number of reactor antineutrinos emitted per one fission in the AZ. We admit that the hardness of the summary $\bar{\nu}_e$ - spectrum at the point $\vec{r}$ equals one unit of hardness if the ratio of densities $F_{\text{Li}}(\vec{r})/F_{\text{AZ}}(\vec{r})$ equals to $1/\bar{n}_\nu$. Then the total spectrum generalized hardness is:

$$H(\vec{r}) = \bar{n}_\nu \frac{F_{\text{Li}}(\vec{r})}{F_{\text{AZ}}(\vec{r})} \quad . \tag{1}$$

This definition is very convenient as in so doing the averaged value generalized hardness of steady spectrum sources (these models are considered in [3,7-9]) is estimated by the value of its productivity factor $k$ of the blanket.

The $\bar{\nu}_e$-cross section in the total spectrum is the additive value of the cross sections in the $\bar{\nu}_e$-flux from AZ and from $^8$Li [14, 15]. In fact the total number of $\bar{\nu}_e$ (crossing the neutrino detector) will be defined the hardness $H$ as:

$$N_{\tilde{\nu}_e} = N_{\text{AZ}} + H(\vec{r}) \times \frac{N_{\text{AZ}}}{\bar{n}}, \tag{2}$$

where $N_{\text{AZ}}$ - number of $\bar{\nu}_e$ from AZ, $\bar{n}_\nu$ - number of $\bar{\nu}_e$ from AZ per one fission, $H(\vec{r})$ - averaged generalized hardness of the total spectrum in the detector position. The second summand determines the number of lithium antineutrinos.

More strictly for density of the total $\bar{\nu}_e$-flux in the point $\vec{r}$ we can write:

$$F_{\tilde{\nu}_e}(\vec{r}) = F_{\text{AZ}}(\vec{r}) + H(\vec{r}) \times \frac{F_{\text{AZ}}(\vec{r})}{\bar{n}}, \tag{3}$$

where $F_{\text{AZ}}(\vec{r})$ - density of the $\bar{\nu}_e$-flux from AZ, $H(\vec{r})$ - generalized hardness in the point $\vec{r}$.

As the cross section is the additive value then for total $\bar{\nu}_e$-spectrum we can write:

$$\sigma_{\bar{\nu}_e p}(\vec{r}) = \sigma^{\text{AZ}}_{\bar{\nu}_e p} + H(\vec{r}) \times \sigma^{\text{Li}}_{\bar{\nu}_e p}. \tag{4}$$

The threshold of the reaction is 1.806 MeV but often (depending on the background) the used threshold is 3 MeV. Taking into account the data of [6] the cross section (4) was calculated as function of the hardness $H$ for the $E_{\text{threshold}}$ = 3, 4, 5, 6 MeV (Fig. 2). At increase of $H$-value the strong rise of the cross section is caused by enlarged part of lithium neutrinos in the total spectrum and energy squared dependence $\sigma_\nu \sim E_\nu^2$. For lithium spectrum the relative yield to the cross section (4) ensured by more high energy neutrinos is significantly larger compare to the AZ spectrum (in calculation we used $\bar{\nu}_e$-spectrum of $^{235}$U [4] as a single fuel isotope). This fact suggest to recalculate the cross section for more higher thresholds. The results (for

$E_{threshold}$ = 4, 5 and 6 MeV) show that for hard total spectrum the lithium yield to the cross section strongly dominates the reactor part [14, 15] (see Fig. 2).

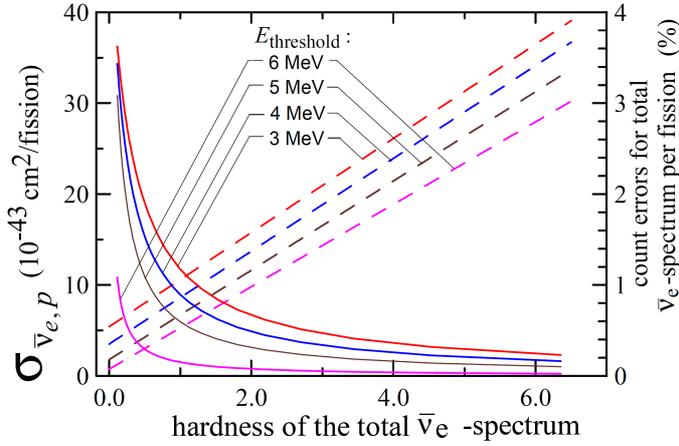

FIG. 2. Cross section of $(\bar{\nu}_e + p \rightarrow n + e^+)$-reaction (dash line) and count errors (solid line; discussion - see below) in the total $\bar{\nu}_e$-spectrum as function of the hardness $H$. Values of cross sections at $H = 0$ correspond to $\bar{\nu}_e$–spectrum from pure $^{235}$U. The results are given for thresholds of registration: 3, 4, 5, 6 MeV.

For the perspective experiments we need to evaluate the advantages from the combined $\bar{\nu}_e$-flux (from AZ plus from $\beta^-$-decay of $^8$Li) given by well known $^8$Li spectrum. The problem of $\bar{\nu}_e$-errors from AZ become very serious as the measurements of $\bar{\nu}_e$-spectrum in Daya Bay, Reno and Double Chooz experiments reveal significant excess of neutrinos with energy 5-7 Mev [1]. This bump in experimental spectrum caused active discussion of the used models, nuclear databases and understanding of results in reactor oscillation experiments.

Here we want to confirm the decrease of count errors in case of such combined $\bar{\nu}_e$-flux. For this purpose we calculated the dependence of errors (in the total $\bar{\nu}_e$-spectrum) on the antineutrino energy and these errors were averaged on their total $\bar{\nu}_e$-spectra for every of thresholds: $E_{threshold}$ = 3, 4, 5, 6 MeV [14-16].

The result dependences of averaged errors on hardness $H$ for the combined spectrum (from AZ with bump in the spectrum plus from $^8$Li) and for specified thresholds are presented in Fig. 2. The accurate calculation with detailed energy bins and evaluation of errors for each bins depending on the hardness $H$ gives sharp decrease of errors for more harder $\bar{\nu}_e$-spectrum. It is confirm the possibility to decrease the count errors in several times compare the significant errors in case of $\bar{\nu}_e$-spectrum of AZ.

It is possible to supply powerful neutrino fluxes of considerably greater hardness by means a facility with a transport mode of operation: lithium is transferred in a closed cycle through a blanket and further toward a remote neutrino detector. In fact we propose to re-pump liquid lithium substance (see below) in continuous conditions (Fig. 3).

For increasing of a flux of hard lithium antineutrinos a being pumped reservoir is constructed in the remote part of the closed loop (in the space close to the $\tilde{\nu}_e$-detector).

Due to the geometrical factor the total $\bar{\nu}_e$-spectrum in the detector volume will be more harder compare to reactor antineutrino spectrum. The closer to the reservoir will be detector then the total spectrum will be harder. Such a facility will ensure not only more hard spectrum in the location of a detector but also an opportunity to investigate $\bar{\nu}_e$-interaction at different spectrum hardness varying a rate of lithium pumping over in the proposed scheme with the closed loop [7, 13,14-17].

The nature lithium consists of two isotope $^6$Li and $^7$Li with concentration 7.5% and 92.5% correspondingly. The beneficial $^7$Li(n,γ)$^8$Li cross section is very small compare to large parasitic absorption on $^6$Li. So, at thermal point (neutron energy $E_n$ ~ 0.025 eV) the cross section $\sigma_\alpha^{thermal}(^6Li) \approx 937$ b, but this one for $^7Li(n,\gamma)$-activation is lower in four orders - $\sigma_{n\gamma}^{thermal} \approx 45$ mb. So we need to ensure high grade of $^7Li$ purification. Results of modeling show that requested $^7Li$ purity must be about 0.9999 (or 0.9998 as minimum level). [3, 7-9, 11-12]. In order to fill the volume of the proposed scheme (Fig. 3) we will need ~22 m$^3$, i.e., 12.2 t of $^7Li$ isotope ($\rho = 0.553$ g/cm$^3$ [18]).

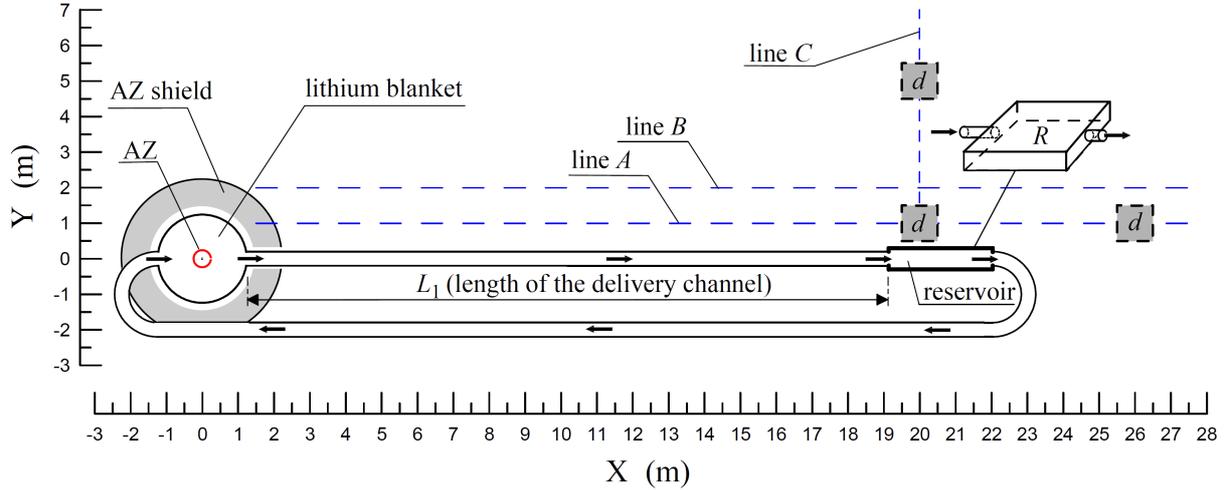

FIG. 3. The principal scheme of $\bar{\nu}_e$-source with variable (regulated and controlled) spectrum for short base line experiment. The all dimensions in the figure are fully correspond to sizes used in the simulation. Liquid lithium substance in the blanket (activated by AZ neutrons) is pumped continuously through the delivery channel to the remote reservoir (volume which is set close to the $\bar{\nu}_e$-detector) and further back to the blanket. The examples of detector positions are labelled as "$d$". The 3-dimensional view of the reservoir (labeled by "$R$") is given separately. At simulation of the antineutrino appearance (see below) we used the coordinates of points along lines $A$ ($y = 1$), $B$ ($y = 2$) and C ($x = 20$). The rate of pumping can be smoothly varied by the installation (not shown) of regime maintenance.

The main input to blanket productivity factor $k$ gives the thermal neutrons. So the logical solution is the next: to use material with high slowing-down power, but with very small absorption in order to ensure significant capture on $^7Li$. We propose to use such perspective substance as a heavy water solution of lithium hydroxides - $^7LiOD$, $^7LiOD \cdot D_2O$ [7-9, 11, 17]. Really this approach will help to solve two problem: 1) to pump a heavy water solution in the scheme with variable spectrum is more simple and safe compare to transport of metallic lithium; 2) the requested mass of high purified $^7Li$ will strongly decrease (the price of installation will be heavy smaller). For our results of simulation (see below) the 22 m$^3$ in volume means that the mass of $^7Li$ in heavy water solution (concentration of LiOD - 9.6 %) will decrease in 18 times and will be 0.71 t. The considered blanket on the basis of heavy water LiOD solution with concentration ~ 9.5% and 0.9999 purity of $^7Li$ possess the productivity $k \geq 0.10$. For smaller LiOD concentration (5.7%) the mass of $^7Li$ will drop in 32.5 times (i. e., 0.39 t) at unimportant decrease of the blanket productivity factor $k$. It is fully realistic to ensure the requested mass of $^7Li$ with purification 0.9999 which widely used for light-water power reactors [19, 20] and permanently produced in significant quantity [21].

We consider the steady state mode of the source operation when fluxes escaping from different parts of the scheme do not vary in time. It was obtained the equation for fluxes of lithium antineutrinos from the blanket and parts of the closed loop [12, 13, 17]. The integral flux of lithium antineutrinos emitted from the pumped volume $V_B$ of the blanket during the time $t$ is:

$$N_B(t) = \lambda_{n\gamma} N_7^0 t \left\{ 1 - [\varphi(V_B)\varphi(V_0 - V_B)] / \lambda_{n\gamma} t_p \varphi(V_0) \right\}, \tag{5}$$

where: $\varphi(y) = 1 - \exp(-\lambda_\beta y/w)$, , $\lambda_{n\gamma}$, $\lambda_\beta$ - rate of (n,γ)-reaction and β⁻-decay; $w$ - volume being pumped over in a time unit (rate of flow); $V_0$ - volume of the whole system, $t_p = V_B/w$ - time of pumping over of the blanket volume; $\lambda_{n\gamma} N_7^0$ is the number of $^8$Li nuclei created in a time unit.

The flux of lithium antineutrinos from a delivery channel during a time $t$ is the next:

$$N_{cd}(t) = \left(\lambda_{n\gamma} N_7^0 t / \lambda_\beta t_p\right) \times \left[ \varphi(V_B) \times \varphi(wt_d) / \varphi(V_0) \right], \tag{6}$$

where $t_d = L_1/\mathcal{V}$ is the time of lithium delivery from the blanket to the pumped reservoir with linear velocity $\mathcal{V}$. The expression (6) allows obtain the flux from different parts of the closed cycle specifying the corresponding time intervals of lithium delivery to the appointed part.

For the simulation we specified the next parameters of the source and regime of the operation. Volume of the compact spherical AZ - corresponds to 51 liters volume of the high flux research reactor PIK constructed now near the Saint-Petersburg in Russia [22, 23]. Thickness of the spherical lithium blanket - 1 m. Volume of the reservoir (rectangular parallelepiped of 0.5 m thickness) was set equal to blanket one. $L_1$ - distance between lithium blanket and pumped reservoir. In the simulation the volume rate of pumping was $w = 2.25$ m³/s. The distance $L_1$ corresponds to the time 1 s of lithium delivery from the blanket to reservoir for appointed $w$ rate.

The several oscillation experiments (LSND [24], SAGE [25], MiniBooNe [26], GALLEX [27], reactor experiments [28] revealed anomalous fluxes and strongly stimulated the discussion on existence of sterile neutrinos and extension of the Standard Model. The considered variants include models with one, two and three type of sterile neutrinos [29 - 32]. Some investigation indicate that squared-mass difference between sterile and active neutrinos - $\Delta m^2 \sim 1$ eV².

The proposed experiment on sterile neutrino search (Fig. 3) is short base line experiment and has advantages namely at short distances where the large hardness is ensured. For short base line setup in case of (3+1)-model of three active neutrinos plus one sterile neutrino the probability of existence at distance $L$ is given by two-flavor model as:

$$P = 1 - \sin^2(2\theta)\sin^2[1.27\Delta m_{41}^2(L(m)/E(MeV)], \tag{7}$$

$$\sin^2(2\theta) = 4|U_{i4}|^2(1 - |U_{i4}|^2), \tag{8}$$

where $\theta$ - angle of mixing, $\Delta m_{41}^2$ (eV²) - maximum squared-mass difference between sterile and active neutrinos (i.e., $|\Delta m^2_{41}| \gg |\Delta m^2_{31}| \gg |\Delta m^2_{21}|$), $U_{i4}$ - element of mixing matrix for active neutrino flavor $i = e, \mu, \tau$.

Probability for (3+2)-model with two sterile neutrinos for short base experiment wiil be:

$$P_e = 1 - 4(1 - |U_{e4}|^2 - |U_{e5}|^2) \times \{|U_{e4}|^2 \sin^2[1.27\Delta m_{41}^2(L/E)]$$
$$+ |U_{e5}|^2 \sin^2[1.27\Delta m_{51}^2(L/E)]\} - 4|U_{e4}|^2|U_{e5}|^2 \sin^2[1.27\Delta m_{54}^2(L/E)]. \tag{9}$$

For simulation of the probability of $\bar{\nu}_e$-existence, fluxes and expected ($\bar{\nu}_e$, $p$)-events in the detector the source volume (Fig. 3) was divided on small cell. Basing on Eq. (5-6) the number of $^8$Li nuclei in any cells was obtained for the given pumping regime. Knowing the reactor and $^8$Li spectrum, the flux from AZ (that also was segmented) and cell fluxes we calculate the flux, spectrum, hardness at the detector positions. The most high level of hardness (that is important for high rate of counts and low errors) is supported in the close space around the voluminous reservoir. For analysis of oscillation we considered the next simple geometry: the detector (or detectors) can be set and shifted along the line A, B or C (Fig. 3). This detector geometry is realistic owing to high count rate in the hard spectrum and possibility to reduce the sensitive volume up to ~ m$^3$ (see below). For AZ spectrum it was taken that the single fuel element is $^{235}$U similar to reactor PIK [22, 23]. The density of $\bar{\nu}_e$ – flux from AZ-is determined by its power $P$ and for distance $R$ is:

$$F\,(\text{cm}^{-2}\cdot\text{s}^{-1}) = \bar{n}_\nu P / 4\pi R^2 \bar{E} \simeq 1.5 \times 10^{12} P\,(\text{MW}) / R^2(\text{m}), \qquad (10)$$

where $\bar{E} \simeq 200$ MeV - mean energy released at $^{235}$U-fission.

In the calculation the applied proton concentration in the detector is typical - about $6.6 \times 10^{22}$ cm$^{-3}$ (as in KamLAND liquid scintillator [33]). The matrix elements for (3+1) and (3+2)-models correspond to best fits of the work [29]: for (3+1) model - $\Delta m^2_{41} = 1.78$ eV$^2$, $U_{e4} = 0.151$; (3+2)a-model - $\Delta m^2_{41} = 0.46$ eV$^2$, $U_{e4} = 0.108$, $\Delta m^2_{51} = 0.89$ eV$^2$, $U_{e5} = 0.124$; (3+2)b-model - $\Delta m^2_{41} = 0.47$ eV$^2$, $U_{e4} = 0.128$, $\Delta m^2_{51} = 0.87$ eV$^2$, $U_{e5} = 0.138$. The update analysis of last neutrino experiments gives some differing global-fit-parameters for sterile neutrinos with $\Delta m^2$ ~1 eV$^2$ [34] compare to Ref. [29].

The calculated errors for count events is given here at 100% efficiency of registration. FIG. 4(a) shows the probability $P$ of existence, hardness and count errors for models (3+1), (3+2)a and (3+2)b at $E_{\text{threshold}} = 3$ MeV for detector positions along lines $A$ (see Fig. 3). Note that hardness $H$ does not depend on the threshold on registration [on definition (1)]. Fig. 5(a) presented the same values for detector positions along line $B$ (see Fig. 3). Fig. 6(a) - shows the above mentioned values for ortogonal line $C$ (see Fig. 3). At coordinates of the reservoir the hardness reaches the maximum with small shift to AZ-position due to decrease of $^8$Li concentration along the flow - see Fig. 4(a) and 5(a). Owing to large lithium mass in the reservoir the maximum of $P$ is detected close to it's position (marked by double arrow) in figures 4(a) and 5(a). Large hardness around the reservoir ensure smallest count errors (below 0.5-1.0%) in the nearby space; shift from line $A$ to line $B$ leads to fall of hardness and increase of errors (compare the Fig. 4(a) and 5(a)). The most rapid decrease of hardness and rise of errors takes place for line $C$ - at the remote from lithium mass (see Fig. 6(a)).

To evaluation the possibility to detect oscillation to sterile neutrinos depending on coordinates let us introduce the functional for opportunity of detecting based on comparison of the maximal $P$ with the current $P(x)$ along $A$-line (see Fig. 3):

$$\Delta_P(x) = [1 - \delta_C(x_{\text{fix}})] \times P(x_{\text{fix}}) - [1 + \delta_C(x)] \times P(x), \qquad (11)$$

where $\delta_C$ - count errors; coordinate $x_{\text{fix}}$ corresponds to maximal $P$ value close to reservoir: $x_{\text{fix}} \simeq 20$ m. The functional helps to search change in probability $P$ avoiding the errors caused by reactor $\tilde{\nu}_e$-spectrum. The positive $\Delta_P$ values determine the $X$ regions where probability of $\bar{\nu}_e$-detecting is higher to level of total spectrum errors.

The $\Delta_P(x)$ results are labelled as (b) and presented in the bottom parts of all Figures 4-6. The analysis for $E_{\text{threshold}} = 3$ MeV along line $A$ (Fig. 4) revealed that the probability to detect oscillation in case of (3+1)-model is close to zero: the $\Delta_P(x)$ curves lay below zero or nearby to it. The model (3+2)a allows to reach the probability up to $\simeq 2\%$, but effects for the model (3+2)b can exceed zero level by 4% (at $x \simeq 6$ m, Fig. 4(b)). The opportunities to reveal the oscillation in the geometries along line $B$ and $C$ are lower (see Fig. 5(b) and 6(b)) that are explained by increased errors for lower hardness in the total spectrum.

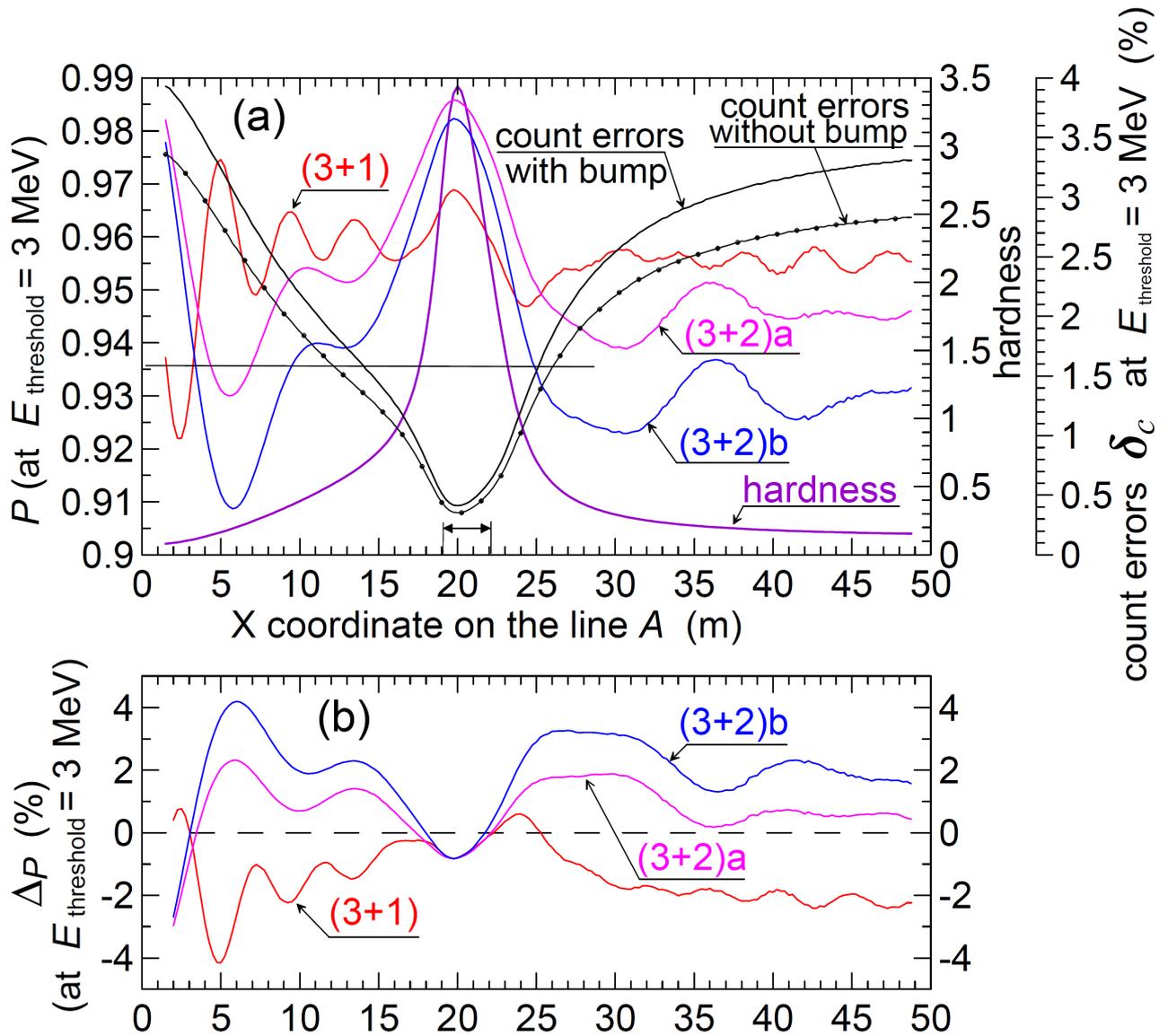

Fig.4. Probability $P$ of $\bar{\nu}_e$-existence for three models [(3+1), (3+2)a and (3+2)b on the part (a)], hardness $H$ of the total $\bar{\nu}_e$-spectrum [part (a)], count errors $\delta_C$ (caused by uncertainties of AZ spectrum) [part (a)] and functional $\Delta_P(x)$ for opportunity of $\bar{\nu}_e$-detecting [part (b) for models: (3+1), (3+2)a and (3+2)b] depending on the $X$ coordinate **along line $A$** (in the detector geometry of Fig. 3). Probability $P$, count errors $\delta_C$ and functional $\Delta_P$ are presented for the threshold of registration $E = 3$ MeV. The all solid lines correspond the values obtained for $\bar{\nu}_e$-spectrum with reactor bump taken into account. The curves with points [count errors $\delta_C$ in the parts (a)] - the errors of $\bar{\nu}_e$-counts for reactor spectrum without bump. Position of the reservoir is shown by the two-sided arrow on the part (a). $X$ coordinates of the positive $\Delta_P$ values are the regions where probability of $\bar{\nu}_e$-detecting is higher to level of total spectrum errors [(see part (b)].

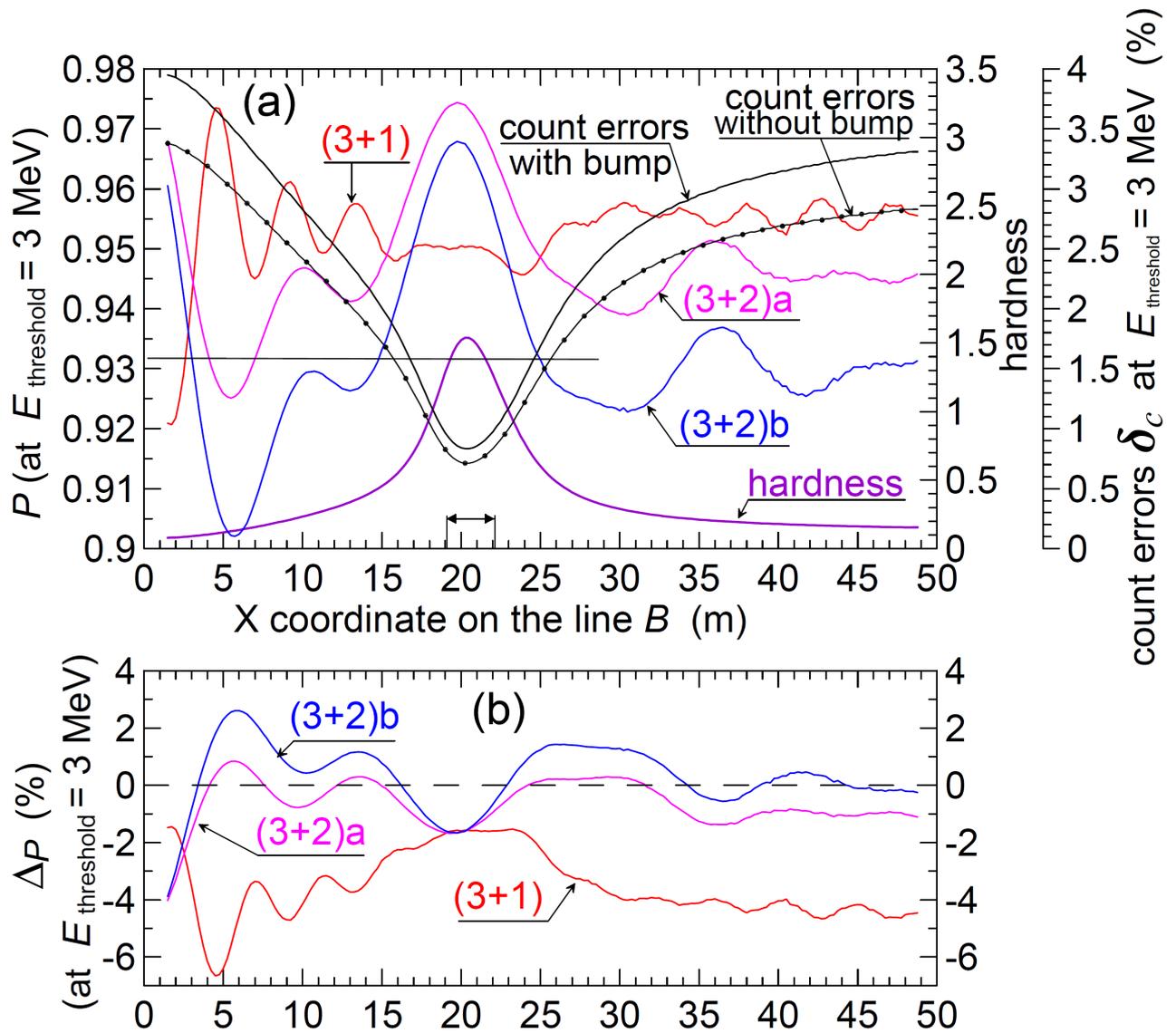

Fig.5. Probability $P$ of $\bar{\nu}_e$-existence for three models [(3+1), (3+2)a and (3+2)b on the part (a)], hardness $H$ of the total $\bar{\nu}_e$-spectrum [part (a)], count errors $\delta_C$ (caused by uncertainties of AZ spectrum) [part (a)] and functional $\Delta_P(x)$ for opportunity of $\bar{\nu}_e$-detecting [part (b) for models: (3+1), (3+2)a and (3+2)b] depending on the $X$ coordinate **along line $B$** (in the detector geometry of Fig. 3). Probability $P$, count errors $\delta_C$ and functional $\Delta_P$ are presented for the threshold of registration $E = 3$ MeV. The all solid lines correspond the values obtained for $\bar{\nu}_e$-spectrum with reactor bump taken into account. The curves with points [count errors $\delta_C$ in the parts (a)] - the errors of $\bar{\nu}_e$-counts for reactor spectrum without bump. Position of the reservoir is shown by the two-sided arrow on the part (a). $X$ coordinates of the positive $\Delta_P$ values are the regions where probability of $\bar{\nu}_e$-detecting is higher to level of total spectrum errors [(see part (b)].

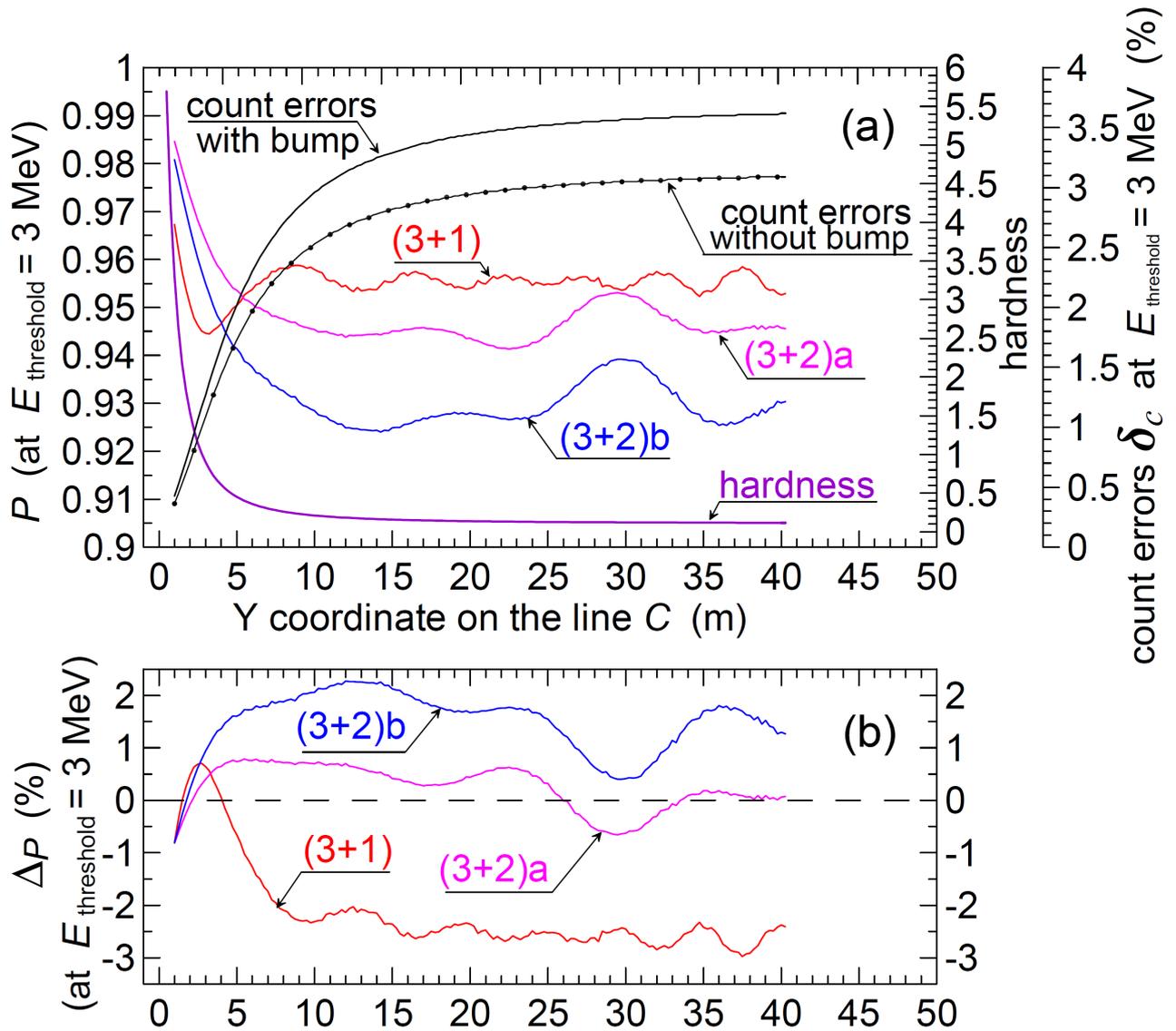

Fig.6. Probability $P$ of $\bar{\nu}_e$-existence for three models [(3+1), (3+2)a and (3+2)b on the part (a)], hardness $H$ of the total $\bar{\nu}_e$-spectrum [part (a)], count errors $\delta_C$ (caused by uncertainties of AZ spectrum) [part (a)] and functional $\Delta_P(x)$ for opportunity of $\bar{\nu}_e$-detecting [part (b) for models: (3+1), (3+2)a and (3+2)b] depending on the $X$ coordinate **along line C** (in the detector geometry of Fig. 3). Probability $P$, count errors $\delta_C$ and functional $\Delta_P$ are presented for the threshold of registration $E = 3$ MeV. The all solid lines correspond the values obtained for $\bar{\nu}_e$-spectrum with reactor bump taken into account. The curves with points [count errors $\delta_C$ in the parts (a)] - the errors of $\bar{\nu}_e$-counts for reactor spectrum without bump. Position of the reservoir is shown by the two-sided arrow on the part (a). $X$ coordinates of the positive $\Delta_P$ values are the regions where probability of $\bar{\nu}_e$-detecting is higher to level of total spectrum errors [(see part (b)].

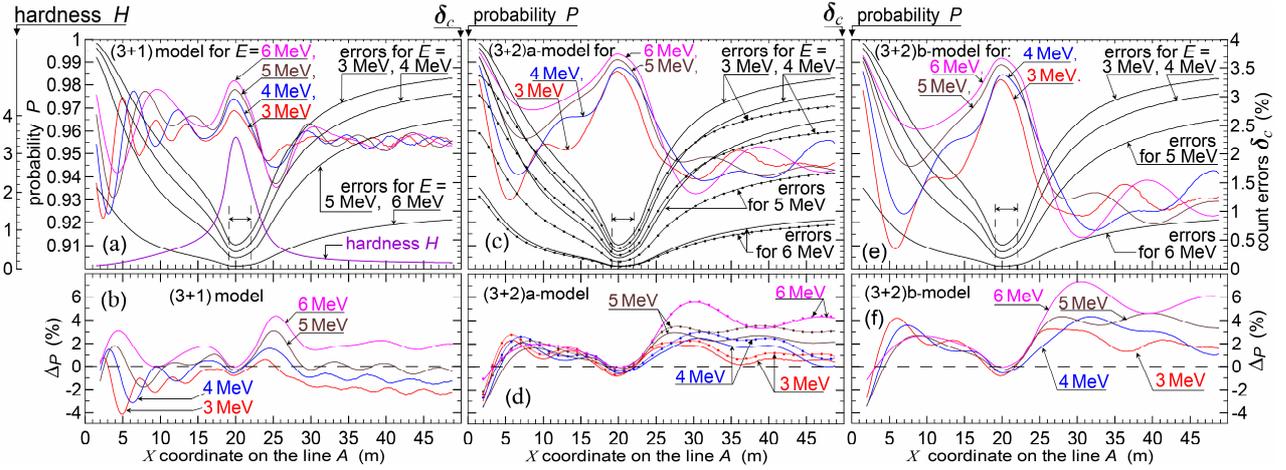

Fig.7. Hardness $H$ of the total $\bar{v}_e$-spectrum (a), probability $P$ of $\bar{v}_e$-existence for three models [(3+1) on the part (a), (3+2)a - part (c) and (3+2)b - part (e)], count errors $\delta_C$ (caused by uncertainties of AZ spectrum) [part (a), (c) and (e)] and functional $\Delta_P(x)$ for opportunity of $\bar{v}_e$-detecting [part (b) for (3+1) model, (d) - for (3+2)a model, (f) - for (3+2)b model] depending on the $X$ coordinate along line $A$. Probability $P$, count errors $\delta_C$ and functional $\Delta_P$ are presented for thresholds of registration $E$ = 3, 4, 5 and 6 MeV (labeled for curves). The all solid lines correspond the values obtained for $\bar{v}_e$-spectrum with bump taken into account. The curves with points [for $\delta_C$ and $\Delta_P$ in the parts (c) and (d)] - the data for model (3+2)a for reactor spectrum without bump. Position of the reservoir is shown by the two-sided arrow on the parts: (a), (c) and (e). $X$-coordinates of the positive $\Delta_P$ values are the regions where probability of $\bar{v}_e$-detecting is higher to level of total spectrum errors.

Before we discussed the advantages of more harder spectrum to decrease the errors (see Fig. 2). The similar analysis of probability oscillation and calculation of $\Delta_P(x)$ values was realized for (3+1), (3+2)a and (3+2)b model in cases of higher thresholds: $E_{threshold}$ = 4, 5, 6 MeV. The obtained data are shown in the Figures 7(b), 7(d) and 7(f). Really at $x$ = 20 m the errors (with bump) decrease in ~ ten times: from 0.4% at $E_{threshold}$ = 3 MeV up to 0.045% at $E_{threshold}$ = 6 MeV. For another $x$ the errors also decrease in several times. As a result in geometry line $A$ for $E_{threshold}$=6 MeV: for (3+1) model $\Delta_P$ > 4% at $x \approx 25$m; for (3+1)a model $\Delta_P$ > 5.5% at $x \approx 30$m; for (3+1)b model $\Delta_P$ > 7% at $x \simeq 30$m. It is interesting to note that significant rize of $\Delta_P$ takes place also at close distances to AZ: for (3+1) model $\Delta_P \simeq 3$% at $x \simeq 4$-5m, $E_{threshold}$ = 6 MeV; for (3+2)a model exists $\Delta_P \simeq$ 1.5-2% at $x \simeq 6-10$m for $E_{threshold}$ = 3-5 MeV; for (3+2)b model $\Delta_P \simeq$ 4% at $x \simeq 6$m for $E_{threshold}$ = 3 MeV and $\Delta_P$ > 3% at $x \approx 7$m for $E_{threshold}$ = 6 MeV. These rize $\Delta_P$ near AZ also can help in testing the model.

But the most reliable detection of $\bar{v}_e$ – disappearance will be at the threshold 6 MeV. So, in order to obtain the more reliable results for (3+1)-model we recommend to work namely at the threshold 6 MeV and compare the counts for x $\simeq$ 20 m and 25 m.

In the Fig. 7(c, d) the errors presented as for reactor spectrum with bump as without bump. The analysis showed the presence of bump can increase the count errors up to $\simeq$ 1% [Fig. 7(d)]. Owing to nuclear reactor (as intensive neutron activator) and remote reservoir (as geometry factor for creation of the hard $\bar{v}_e$ – spectrum) the proposed source ensure high neutrino flux in the space close to reservoir. Here the $^8$Li isotope acts as effective shifter in forming of the hard spectrum. The dependence of cross section as $\sigma_\nu \sim E_\nu^2$ strongly amplifies the effect increasing the number of ($\bar{\nu}_e$,p)-events in the detector. The results of calculated fluxes for points on the line $A$ (see detector geometry in Fig. 3) is given at the Fig. 8. Thank to the reservoir the total flux has "lithium bump" close to the reservoir position. The lithium flux strongly dominate over the reactor flux in the wide $x$-interval.

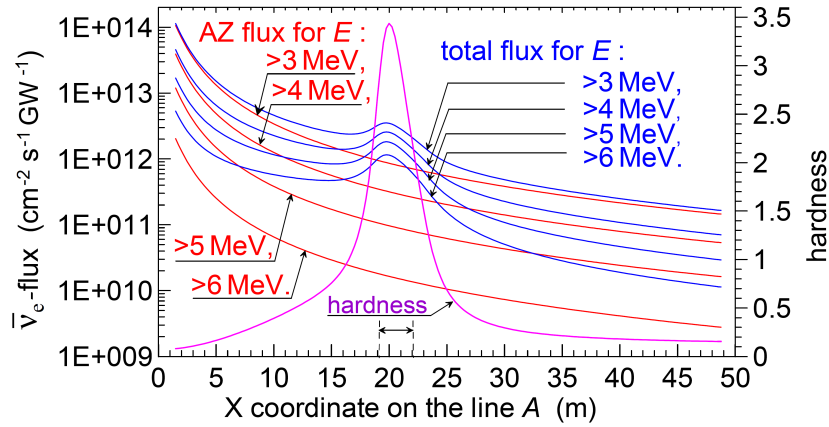

Fig. 8. The densities of $\bar{\nu}_e$-fluxes [from antineutrinos of active zone (AZ) and from whole mass of $^8$Li in the installation] and total-$\bar{\nu}_e$-spectrum-hardness $H$ depending on X-coordinate along line $A$ (the geometry of Fig.3)]. The fluxes are presented for different thresholds of registration ($E_{threshold}$ = 3, 4, 5 and 6 MeV). The all results are normalized on the reactor power 1 GW

Intensive hard neutrino flux gives high yield in $\bar{\nu}_e$-events in detector volume. The expected number of detector events as function of coordinates (line $A$ geometry) is presented in the Fig. 9.

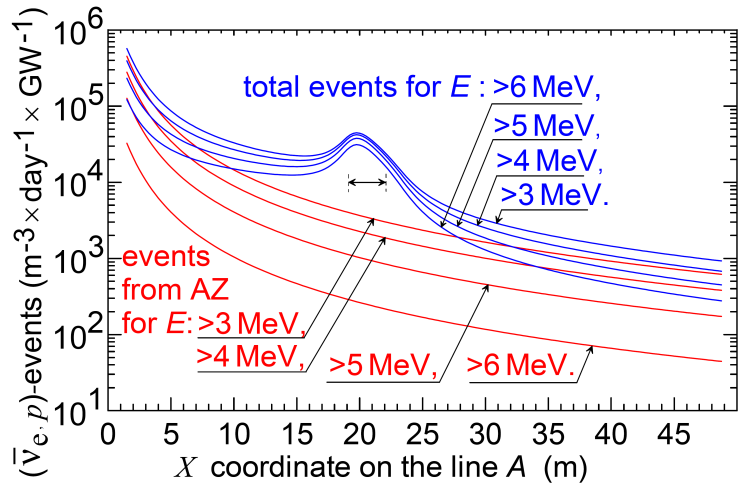

Fig. 9. The expected normalized number of ($\bar{\nu}_e$,p)-events in the detector depending on the $X$ coordinates along the line $A$ (geometry of the Fig. 3). The results are given for thresholds of registration 3, 4, 5, and 6 MeV. The total number of events is ensured by $\bar{\nu}_e$ from $^8$Li and AZ. Position of the reservoir is shown by the two-sided arrow.

The dependencies are calculated for thresholds of registration from 3 to 6 MeV. The results are normalized per cubic meter, day and gigawatt of the reactor power. Note that really the power of the investigating reactors is lower; the exclusively interesting variant (for the discussed $\bar{\nu}_e$-source) of the such reactor is now constructed PIK with the thermal power 100 MW [22,23]. The events ensured by the $^8$Li antineutrinos are strongly dominates owing to the hardness of $\bar{\nu}_e$ – spectrum of $^8$Li. Compare to reactor yield the $^8$Li ensure strong rise of registration rate.

The purpose of the work is to investigate the possibility for sterile neutrino search on the base of the proposed intensive $\bar{\nu}_e$-source with hard spectrum. The idea of the source is (n,γ)-activation of pure $^7$Li near reactor active zone and lithium transfer to remote detector by the loop scheme. For realization the more perspective way is use the heavy water solution of $^7$Li instead of metallic lithium: the requested mass of purified $^7$Li decrease in 18 times and will be 0.71 t. Today the requested mass of such pure $^7$Li exist and can received from the known producers.

The total $\bar{\nu}_e$-spectrum is created by reactor one (fast decreasing one and known with significant errors) and well known hard $^8$Li spectrum. Owing to cross section dependence $\sigma_\nu \sim E_\nu^2$ the number of neutrino interaction strongly increases at rise of the hardness (thank to $^8$Li neutrinos) in the total spectrum. The definition of the hardness $H$ was introduced and dependence of $(\bar{\nu}_e,p)$– cross section from $H$ value was obtained. The second very important feature from addition of $^8$Li neutrino flux is large decrease of count errors for more harder total spectrum. The function of errors from $H$-value was obtained.

It was considered the variant of the $\bar{\nu}_e$ source with realistic dimension and regime of operation. The possible scheme of the experiment for seach of sterile neutrinos with $\Delta m^2$ ~1 eV$^2$ is discussed for schemes (3+1) and (3+2). For this cases the total antineutrino fluxes were calculated taking into account both lithium and reactor spectra and corresponding errors, the dynamics of lithium transfer and dimensions of all parts of the installation. The oscillation probabilities for (3+1)-model and two variants of (3+2)-models were simulated for wide intervals of detector positions. For possible detector positions is was proposed the scheme and calculated the coordinates for search $\bar{\nu}_e$-disappearance outside the spectrum errors. High rate of the detector counts allows to use compact neutrino detector (~ m$^3$).

The author thankful to Yu. S. Lutostansky for helpful discussion. The author wish to express my full appreciative to L. B. Bezrukov, B. K. Lubsandorzhiev, D. K. Nadezhin, and I. I. Tkachev for support and large interest to the work.